\RequirePackage{lineno}
\documentclass[aps, prl,
floatfix,twocolumn,superscriptaddress,
groupedaddress
]{revtex4-1}
\pdfoutput=1
\usepackage{t1enc}
\usepackage[utf8]{inputenc}
\usepackage[T1]{fontenc}
\usepackage{lmodern}
\usepackage{graphicx}
\usepackage{epsfig}
\usepackage{amssymb}
\usepackage{dcolumn}% Align table columns on decimal point
\usepackage{bm}% bold math
\usepackage{color}
\usepackage{float}
\usepackage{hyperref}
\usepackage{natbib}
\usepackage{footnote}
\usepackage{ulem}
\usepackage{hhline}
\newcommand{\BFAP}{\ensuremath{\mathrm{BaFe}_{2}{\mathrm{(As}_{1-x}\mathrm{P}_{x}}\mathrm{)}_2}}
\newcommand{\EFAP}{\ensuremath{\mathrm{EuFe}_{2}{\mathrm{(As}_{1-x}\mathrm{P}_{x}}\mathrm{)}_2}}

\begin{document}

\title{Evidence of hot and cold spots on the Fermi surface of LiFeAs}

\author{J.\,Fink,$^{1,2,3}$ J.\,Nayak$,^2$ E.D.L.\,Rienks$,^{1,3}$ J.\,Bannies$,^2$ S.\,Wurmehl$,^{1,3}$ S.\,Aswartham$,^1$ I.\,Morozov$,^1$ R.\, Kappenberger$,^1$ 
 M.A.\,ElGhazali,$^{2,3}$ L.\,Craco,$^{1,4}$ H.\,Rosner,$^2$ C.\,Felser,$^2$  and B.\,B\"uchner$^{1,3}$}
\affiliation{
\mbox{$^1$Leibniz Institute for Solid State and Materials Research  Dresden, Helmholtzstr. 20, D-01069 Dresden, Germany}\\
\mbox{$^2$Max Planck Institute for Chemical Physics of Solids, D-01187 Dresden, Germany}\\
\mbox{$^3$Institut f\"ur Festk\"orperphysik,  Technische Universität Dresden, D-01062 Dresden, Germany}\\
\mbox{$^4$Instituto de Fisica, Universidade Federal de Mato Grosso, 78060-900 Cuiabá, MT, Brazil}\\
}

\date{\today}

\begin{abstract}\
 Angle-resolved photoemission spectroscopy (ARPES) is used to study  the  energy and momentum dependence of the inelastic scattering rates and the mass renormalization of charge carriers in LiFeAs at several high symmetry points in the Brillouin zone. A strong and  linear-in-energy scattering rate is observed  for sections of the Fermi surface  having predominantly Fe $3d_{xy/yz}$ orbital character on the inner hole  and on electron pockets. We assign them to hot spots with marginal Fermi liquid character inducing high  antiferromagnetic and pairing susceptibilities. The outer hole pocket, with Fe $3d_{xy}$ orbital character, has a reduced but still linear in energy scattering rate. Finally, we assign sections on  the middle hole pockets with  Fe $3d_{xz,yz}$ orbital character and on the electron pockets with  Fe $3d_{xy}$ orbital character  to cold spots because there we observe a quadratic-in-energy scattering rate with  Fermi-liquid behavior. These cold spots  prevail the transport properties. Our results indicate a strong $\it{momentum}$ dependence of the scattering rates. We also have indications that  the scattering rates in correlated systems are fundamentally different from those in non-correlated materials because in the former the Pauli principle is not operative. We compare our results for the scattering rates with combined density functional plus dynamical mean-field theory  calculations.   The work provides a generic microscopic understanding of macroscopic properties of multiorbital unconventional superconductors.
\end{abstract}

%\pacs{74.25.Jb, 74.70.Xa, 79.60.-i}

\maketitle

\paragraph{Introduction.} Since the discovery of iron-based superconductors (FeSCs)\,\cite{Kamihara2008} there is  an ongoing debate about  whether their electronic structure is more itinerant or localized\,\cite{Johnston2010}. Transport properties and optical spectroscopy indicate predominantly  a Fermi liquid behavior~\cite{Rullier-Albenque2009,Barisic2010,Tytarenko2015}. On the other hand, high-$T_c$ $s^{\pm}$ superconductivity and antiferromagnetism are believed to occur due to strong correlation enhanced scattering of the charge carriers between hole  and electron pockets~\cite{Mazin2008a}. In these multi-orbital systems, it is likely that the different properties are related to the  intrinsic  multi-orbital nature of electron and hole pockets. This is supported by combined density functional plus dynamical mean field theory (DFT+DMFT) calculations, which pointed out that the strength of correlation effects strongly depends on the orbital character of the bands due to their different  respective filling \,\cite{Haule2009,Aichhorn2009,Aichhorn2010,Medici2011,Werner2012,Razzoli2015,Bascones2016} and that for half-filled orbitals, correlation effects are not only determined by the onsite Coulomb interaction but also  by  Hund's exchange interaction. 

The strength of correlation effects, however, may be  determined not only by the orbital character of the bands but also by the nesting conditions and, thus, by the momentum\,\cite{Graser2009,Kemper2011}. In order to obtain a microscopic understanding of these effects, momentum dependent information on the character of the states near the Fermi level ($E_F$) is necessary. Angle-resolved photoemission spectroscopy (ARPES) is a suitable method to obtain this momentum dependent information because it provides the energy ($E$) and momentum ($\mathbf{k}$) dependent scattering rates $\Gamma$ or lifetimes $\tau=\hbar/\Gamma$ of the charge carriers\,\cite{Damascelli2003} which are related to the imaginary part of the self-energy $\Im\Sigma$ by $\Gamma=-2Z\Im\Sigma$, where  $Z=m_b/m^*=\frac{1}{1-\frac{\partial\Re\Sigma}{\partial E}}$ is the quasiparticle residue and $m^*/m_b$ is the mass renormalization\,\cite{Mahan2000}. 
 The latter is derived from a comparison of the ARPES data with DFT calculations. 
Here, we use ARPES  to study the scattering rate   of superconducting LiFeAs\,\cite{Tapp2008} in its normal state. This  tetragonal compound without  dopant atoms to induce disorder, is particularly suited for ARPES studies\,\cite{Lankau2010,Nascimento2009,Heumen2011}. 

\begin{figure*}[tb]
\centering
 \vspace{-2.5 cm}
\includegraphics[angle=0,width=1.0\linewidth]{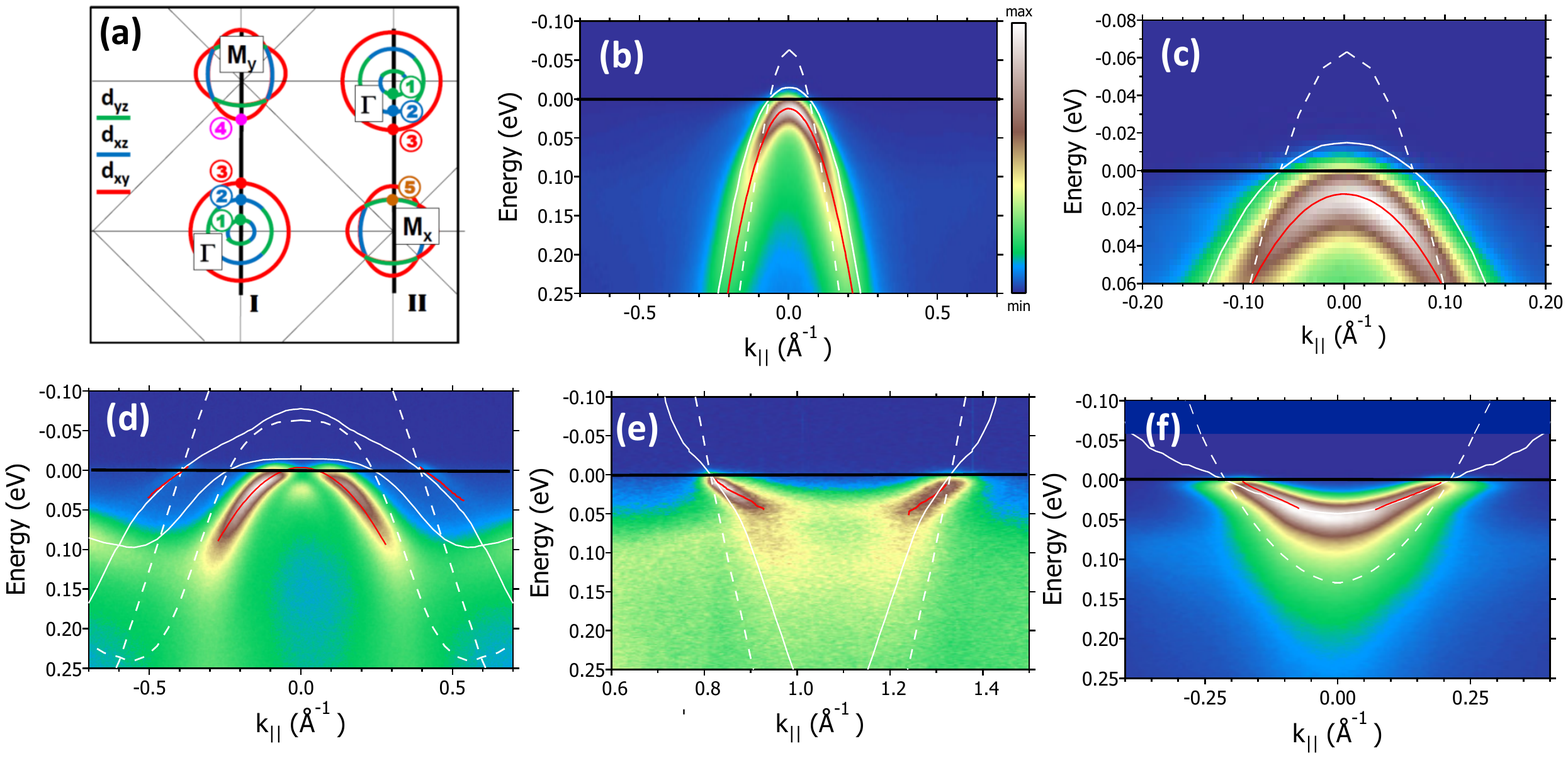}
\vspace{-2 cm}
 \caption{
(a) Schematic Fermi surface of LiFeAs in the $k_z=0$ plane. Sections with different orbital character are marked by different colors\,\cite{Graser2009,Kemper2011}. (b) - (f) Energy momentum distribution maps (EMDM) measured at various points indicated in (a). Red lines: dispersions derived from fits of ARPES data. Dashed white lines: dispersions from DFT calculations. Solid white lines: dispersions from DFT+DMFT calculations.  (b) Data from the  inner hole pocket with $d_{yz}$ orbital character,  measured near  $\Gamma$ [point (1)] along  cut I   with $s$-polarized photons (energy $h\nu$ = 82  eV). (c) Zoom into the top of the inner hole pocket.  (d) Similar data as in (b) but measured with $p$-polarized light (energy $h\nu$ = 49  eV). The outer [$d_{xy}$ orbital character, point (3)] and the middle [$d_{xz}$ orbital character, point (2)] hole pockets are visible. (e) EMDM measured along cut I near $M_y$ with $s$-polarized photons ($h\nu$ = 123 eV)  showing the outer electron pocket [$d_{xy}$ orbital character, point (4)]. (f) EMDM  measured with $s$-polarized photons ($h\nu$ = 84 eV) along cut II showing the  electron pocket [$d_{yz}$ orbital character, point (5)]  near the $M_x$ point.  
}
\label{edm}
\centering
\end{figure*}

 There are numerous ARPES studies on LiFeAs\,\cite{Borisenko2010,Kordyuk2011,Umezawa2012,Lee2012,Miao2016,Brouet2016,Hajiri2016,Day2018}. In the investigations of the scattering rates, one of the main results was that  $\Gamma$ depends on the orbital character rather than on the position on the Fermi surface. This result was possibly biased by DFT+DMFT calculations on FeSCs\,\cite{Yin2011a,Ferber2012} which are based on a local, not  momentum dependent approximation for the correlation effects. The present high resolution ARPES data together with a new evaluation method comes to a different  conclusion: the inelastic scattering rates depend predominantly on the momentum and only indirectly on the orbital character.  Our study  yields information on the location of hot (large $\Gamma$) and cold (small $\Gamma$) spots within the Brillouin zone (BZ). This will improve our microscopic understanding of  the magnetic and superconducting susceptibilities (determined by the hot spots) and normal state transport properties (determined by the cold spots).

\paragraph{Experimental.} 
 LiFeAs ($T_c=17$ K) single crystals were grown  using the self-flux technique\,\cite{Morozov2010}.
ARPES measurements were conducted at the $1^2$ and $1^3$-ARPES end stations attached to the beamline UE112 PGM at BESSY, equipped with Scienta R8000 and Scienta R4000 energy analyzers, respectively. All data presented in this contribution were taken in the normal state at temperatures between 20 and 35 K.  The achieved energy and angle resolutions were between 4 and 15 meV and 0.2$^\circ$, respectively. Polarized photons with energies $h\nu=20-130$ eV were employed to reach different $k_z$ values in the BZ and spectral weight with a specific orbital character\,\cite{Fink2009,Moser2017}. An inner potential of 12 eV was used to calculate the $k_z$ values from the photon energy~\cite{Borisenko2015a}. 

\paragraph{Data evaluation and electronic structure calculations.}
Usually, in ARPES the  scattering rates are derived from the width of the spectral weight at constant energy\,\cite{Damascelli2003,Valla1999}. The lifetime broadening in energy space is obtained by multiplying the momentum width by the velocity. This  works well for a linear dispersion because in  this case the velocity $v$ is constant in energy  and therefore the contribution from the elastic scattering is constant. It is also possible to evaluate the  inelastic lifetime broadening  from a parabolic dispersion but then one has to take into account the energy  dependence of the velocity. At the top and the bottom the velocity is zero and therefore this method does not work at these energies. In the present work we use a multivariate fit of the measured data which is superior to the methods  described above. The finite energy and momentum resolution was taken into account by convolutions with Gaussian functions. To obtain the inelastic scattering rates due to electron-electron interaction we subtracted the energy dependent elastic scattering rates $\Gamma_e(E)$. The later is calculated from the relation $\Gamma_e(E)=v(E)w_0$. The velocity $v(E)$ is taken from the ARPES dispersion. The momentum width $w_0$ at $E_F$, corresponding to the inverse mean free path, can be derived using $\Gamma_t(0)=\Gamma_e(0)=v(0)w_0$. This follows from the fact that the inelastic scattering rate  $\Gamma_i(0)$ is zero at $E_F$\,\footnote{see information presented in the Supplement}.

 We performed density functional band structure calculations
within the local  density approximation including spin orbit coupling, 
using  experimental structural parameters~\cite{Tapp2008} and 
the full-potential local-orbital code FPLO~
\cite{Koepernik1999} 
(version fplo18.00-52) with the Perdew-Wang exchange correlation
potential~\cite{Perdew1996}.
Similar to earlier DFT+DMFT studies~\cite{Craco2008,Craco2017},
the local self-energies due to the correlated Fe $3d$ problem, intrinsic 
to Fe-based superconductors are obtained with the multi-orbital iterated 
perturbation theory  as impurity solver~\cite{Craco2008a}, where 
$U=2.5$~eV and $J_H=0.7$~eV are used.

From a parabolic fit to the ARPES data and to the DFT results close to $E_F$ we derive the renormalized mass $m^*$ and the bare particle mass $m_b$, respectively. The ratio $m^*/m_b$ yields the mass renormalizations.

\paragraph{Results.}
In Fig.~1\,(a) we present a schematic Fermi surface of typical ferropnictides in a selected region in reciprocal space in the  $k_z=0$ plane. In LiFeAs for $k_z$=0, no inner hole pocket is visible in the Fermi surface because it is about 12 meV below $E_F$. Using thick black solid lines we mark  two cuts (I and II) along which we have performed ARPES measurements.  In Fig.\,1\,(b)-(f) we show energy-momentum distribution maps of  hole and electron pockets, recorded along cut I and cut II  using different photon polarizations  to select bands with different orbital character\,\cite{Fink2009}. To demonstrate the existence of spectral weight with $yz$ orbital character at $E_F$ due to a correlation induced broadening of the band, we show a zoom in of Fig.~1(b) in Fig.~1(c). In Fig.~1(b)-(f) we have added the dispersions calculated by DFT and DFT+DMFT.
Using other  photon energies, we have collected analogous data of the spectral weight in planes corresponding to  $k_z=\pi/c$. 
\begin{figure}[tb]
\centering
\vspace{-2cm}
\hspace{-1.5cm}
 \includegraphics[angle=90,width=10cm]{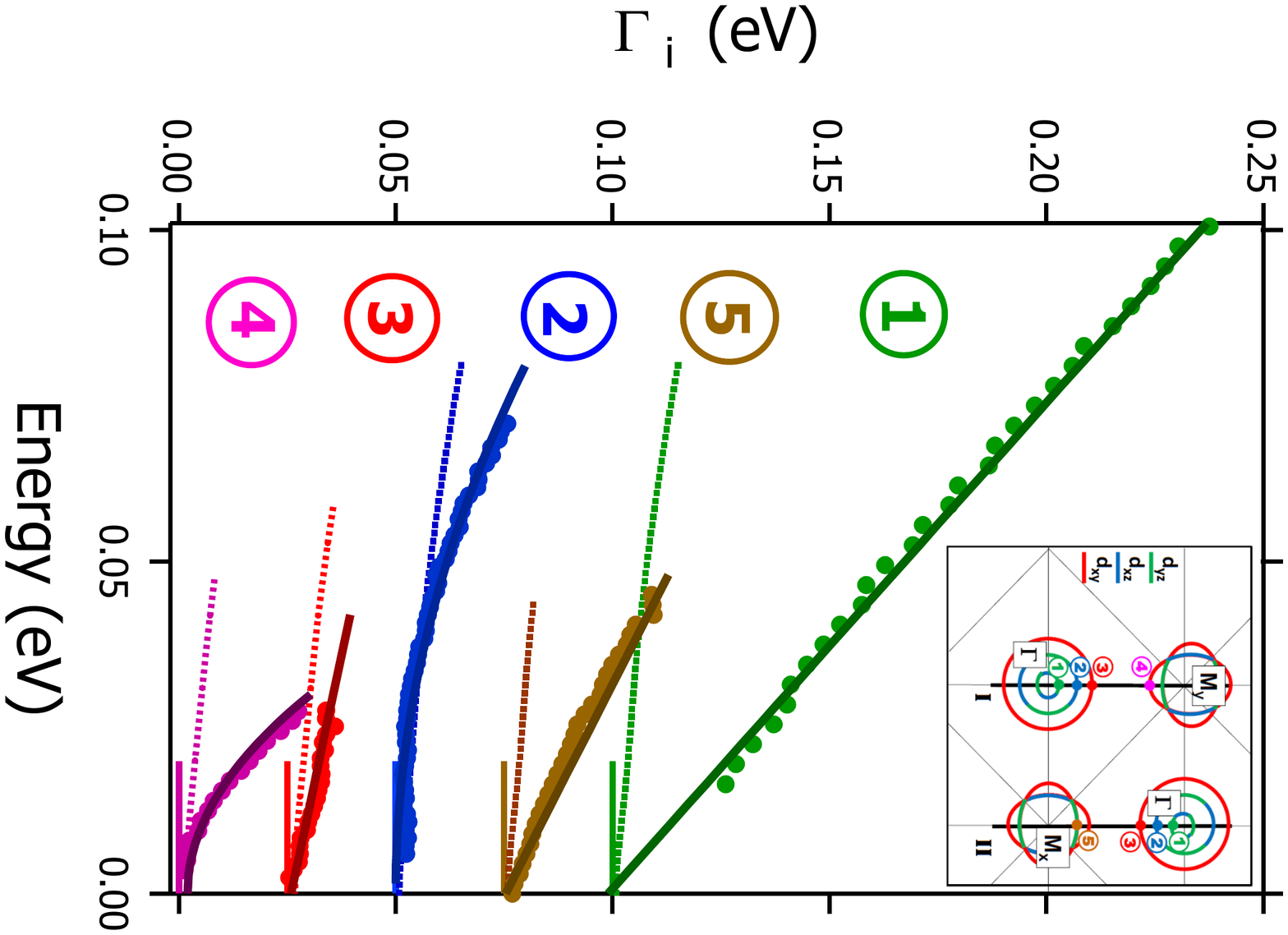}
\vspace{-1cm}
\caption{
Insert: the same as Fig~1(a). Lower panel: lifetime broadening $\Gamma_i$  (large dots) for  various points  marked in the insert.  Solid lines: fit to the data points. Small dots: DFT+DMFT calculations of the scattering rates due to multi-orbital electron-electron interactions. The curves are plotted in a stacking mode. The vertical shift between the curves is 0.025~eV. Colors are related to  particular sections of the Fermi suface (see insert).  
} 
\label{mass}
\centering
\end{figure}

In Fig.~2 we compare $\Gamma_i(E)$ from  ARPES with  DFT+DMFT results. As shown previously for the inner and outer hole pocket in other ferropnictides~\cite{Fink2017}, for all points the theoretical values are considerably smaller than the experimental  data.
In the analyzed energy regions, the data for points (1), (3), and (5) can be well  fitted by a linear relationship $\Gamma=\beta E$. At points (2) and (4) $\Gamma_i$ can be fitted by a quadratic relationship $\Gamma=\gamma E^2$.   We have not found a clear indication of  significant electron-phonon coupling which would result in a step in $\Gamma_i(E)$ and a kink in the dispersion near the Debye energy of $\approx 0.03$ eV\,\cite{Engelsberg1963}.  In the case when a band does not cross $E_F$, e.g. the top of the inner hole pocket is $\approx 0.012$ eV below $E_F$,  the data are limited  at low energies.  At high energies the data are limited by a finite band width,  overlapping  bands, or in the case when the spectral weight can no longer  be distinguished from the background.
\begin{table*}[t]
\centering
\caption{ARPES data of LiFeAs derived at various high-symmetry points (HSP) [Fig. 1 (a)] for $k_z=0$ and $k_z=\pi/c$ having predominantly an orbital character OC. $ m^*/m_b$ is the mass renormalization. $w_0$ is the momentum width near $E_F$ caused by elastic scattering. $\beta$ gives the slope in regions of a linear energy dependence of the scattering rate for points (1), (3), and (5).  The parameter $\gamma$ is related to the quadratic increase of the scattering rates at points (2) and (4).  n.a. means not applicable.}
\label{tab:1}    
%
% For LaTeX tables use
%
\begin{ruledtabular}
\begin{tabular}{ l l l l l l l l l  l }
%\hhline { = = = = = = = = = =}
HSP & OC & $m^*/m_b$   & $w_0
(\text{\AA}^{-1})$ & $\beta$ & $\gamma$ (eV)$^{-1}$  & $m^*/m_b$   & $w_0(\text{\AA}^{-1})$ & $\beta$ & $\gamma$ (eV)$^{-1}$    \\  
\hline\noalign{\smallskip}
%\cline{1-10}\noalign{\smallskip}\\
     &        & $k_z=0$       &                         &                  &                  & $k_z=\pi/c$ &                 &           &                       \\
\cline{3-10}
(1) & $yz$ & 1.6$\pm$0.4 & 0.00$\pm$0.01 &1.3$\pm$0.1 &n.a.          &n.a.           &n.a.             &n.a.       &n.a.          \\ 
(2 )& $xz$ & 1.3$\pm$0.5 & 0.03$\pm$0.1  &n.a.         &5$\pm$ 4 & 2$\pm$0.2 & 0.03$\pm$0.1&n.a.       &18 $\pm$4     \\
(3) & $xy$ &3.9$\pm$0.1 & 0.03$\pm$0.01 &0.3$\pm$0.1 &n.a.             &3.6$\pm$0.1  & 0.05$\pm$0.03 &0.7$\pm$0.2 &n.a              \\ 
(4) & $xy$ & 4.8$\pm$0.2 & 0.05$\pm$0.01 &n.a.        &32$\pm$5   &3$\pm$1& 0.01 $\pm$0.02&n.a.   &25$\pm$5\\
(5) & $yz$ & 2.3$\pm$0.1 & 0.05$\pm$0.01 &0.7$\pm$0.1 &n.a.      &1.8$\pm$0.1 & 0.10$\pm$0.02 & 0.6$\pm$0.2&n.a.          \\
\end{tabular}
\end{ruledtabular}
\end{table*} 

The parameters describing the energy dependence of the inelastic scattering rates ( $\beta$ and  $\gamma$) are collected in Table I for data corresponding to $k_z=0$ and $k_z=\pi/c$. In addition we present  values of $w_0$ which is used to subtract the contributions from $\Gamma_e$ (see above and Supplement \,\footnote{see information presented in the Supplement}).  The error bars are estimated from the analysis of data taken at different photon energies and different Brillouin zones.

Moreover we present in Table I  the  derived mass renormalizations near $E_F$. These mass renormalizations slightly decreases with increasing binding energy (not shown).   Within error bars there is no $k_z$ dependence of all values presented in Table I.

\paragraph{Discussion.} Comparing the experimental, the DFT, and the DFT+DMFT dispersions (see Fig.~1), in all cases the mass renormalization near $E_F$ is well described by DFT+ DMFT calculations. Also no shift between the ARPES data and the DFT+DMFT results is observed at points (3), (4), and (5). However, near $k=0$ and $k_F$,  shifts of about 0.07 eV to higher energies are observed at points (1) and (2)  between the DFT dispersion and the ARPES dispersion. This shift leads to a shrinking or disappearance of the middle and the inner hole pockets, respectively. This shift is not reduced near the calculated $k_F$ points by DFT+DMFT calculations (see also Ref.~\cite{Ferber2012,Borisenko2015}).

 Regions with a linear increase of the scattering rates as a function of energy have been  detected in other FeSCs and related compounds and were discussed in previous publications \,\cite{Fink2015,Fink2017,Avigo2017,
Nayak2017} in terms of momentum and not orbital dependent  strong correlation effects and the proximity to the Planckian limit~\cite{Zaanen2004}.

Interestingly, the scattering rate at point (1) extrapolates to zero and not to the top of the band at $E_t \approx 0.012$ eV. The latter is expected for a normal Fermi liquid behavior because interband transitions between the inner hole pocket   and the inner electron pocket are not allowed for energies less than $E_t$ because of the Pauli principle. This was also discussed for the superconducting case where the  scattering rate should go to zero at three times  the superconducting gap energy $\Delta$~\cite{NormanDing1998}. However, in highly correlated systems in which the Fermi edge in momentum space is broadened, the Pauli principle is no more operative~\cite{Fink2016} and therefore the scattering rate should extrapolate to zero with a finite lifetime broadening at $E_t$. The experimental observation that $\Gamma_i(E)$ extrapolates to zero and not to  $E_t$  is an important result indicating that interband transitions and their related properties such as magnetic or pair susceptibilities are fundamentally different between correlated and uncorrelated systems. In this context, we mention recent ARPES data on Ni in which also an extrapolation of the linear-in-energy scattering rate to zero was observed for the gapped majority Ni $3d$ band\,\cite{Sanchez-Barriga2018}. A detailed discussion of the interband transition is presented in the Supplement\,\footnote{see information presented in the Supplement}.
  
Different from our previous studies on \BFAP\    and \EFAP\ \cite{Fink2015} in which all energy dependencies of the inelastic scattering rates were assumed to be linear, we observe in the present investigation for  the middle hole pocket and for $d_{xy}$ sections on the electron  pockets  [point (4)] a Fermi liquid behavior, i.e., a quadratic increase as a function of energy , in agreement with DFT+DMFT:  $\Gamma_i=\gamma E^2$ (Fig.~2).\ This  indicates coherent electronic states at this point at low energies.  The prefactor $\gamma \approx30$~eV$^{-1}$ of this energy dependence at point (4) is very  large which leads to a crossover to incoherent states at rather low energies of $\approx 0.03$ eV corresponding to room temperature.
We contrast the  large prefactor $\gamma \approx30$~eV$^{-1}$  detected at this point with a much smaller prefactor $\gamma \approx0.28$ eV$^{-1}$ derived for a weakly correlated electronic structure of a Mo surface~\cite{Valla1999}. Regarding the steep energy dependence, an orbital selective crossover has been detected as a function of temperature in ARPES measurements of FeSCs~\cite{Yi2013,Yi2015,Miao2016}. 

Naively one would expect that because of the connection of  $\Re\Sigma$ with $\Im\Sigma$ by the Kramers-Kronig relation, the mass renormalization $m^*/m_b$ would scale with the strength of the scattering rates. Looking at Table I to the data at  point (1) and (3)  the ratio between $m^*/m_b$ and  $\beta$ is about 0.4 and 3, respectively. Regarding the relationship between $\Gamma$, $Z$, $m^*/m_b$, and $\Im\Sigma$ presented in the Introduction, a small scattering rate together with a small $Z$ (large $m^*/m_b$) yields a large $\Im\Sigma$ and thus via the Kramers-Kronig relation a large mass renormalization. This shows that also the data for $\Gamma$ and $m^*/m_b$ at points (1) and (3) may be compatible with the Kramers-Kronig relation. 

We point out that  the location of the hot spots coincides with regions where the highest superconducting gaps were detected~\cite{Borisenko2012,Umezawa2012}.  In particular we mention that the inner $yz$ hole is still important for superconductivity because due to correlation induced broadening of the bands there is still spectral weight at $E_F$ [see Fig.~1(c)]. The reason for this is that  the difference between $E_F$  and the maximum of the band ($E_t\approx0.012$ eV) is smaller than the coupling energies of spin fluctuation (of the order of 0.01 to 0.1 eV~\cite{Johnston2010})
and should therefore, in an Eliashberg model, contribute to the superconducting transition temperature~\cite{Pickett1982}. The present discussion is also important for many other iron-based superconductors, e.g. the ferrochalcogenides, where the top of the hole pockets are very close to $E_F$.

The observation of Fermi liquid behavior on hole and  electron pockets is in line 
 with transport data~\cite{Rullier-Albenque2012} which derived a  Fermi liquid behavior both in hole and electron pockets. 
 Quantum oscillation experiments~\cite{Coldea2009,Putzke2012}  came to the conclusion that the scattering rates for electrons are smaller than those of holes in agreement with the present result. Their derived mass enhancements are comparable to the present results.
 % Neutron scattering results
%~\cite{} 
%are partially in agreement with our experiments. Ref.~
%\cite{}
%come to the conclusion that the highest susceptibilities are related to $yz$ orbitals which supports our results on the scattering rates. On the other hand Ref.
%~cite{}
% derive the strongest contributions are related to $xy$ orbitals, which in our experiment do not show the highest scattering rates. Here we point out that the suceptibilities do not directly determine the scattering rates. 
Regarding the quadratic increase of $\Gamma_i$ at points (2) and (4) we mention that our data are also consistent with optical spectroscopy data~\cite{Barisic2010,Tytarenko2015}  because the optical conductivity in a multi-orbital system is dominated by cold spots.

\begin{acknowledgments}
This work has been supported by the Deutsche Forschungsgemeinschaft (DFG) through the Priority Programme SPP1458, through the Emmy Noether Programm in project WU595/3-3 (S.W.) and  through Research Training Group GRK 1621. L.C.'s work is supported by CNPq (Grant No. 304035/2017-3). M.E.’s work is supported by the DFG Project Number GRK 1621. 
\end{acknowledgments}

\bibliographystyle{apsrev4-1}
\bibliography{Pnictide}

\end{document}

\title{Supplementary material for: Evidence of hot and cold spots on the Fermi surface of LiFeAs}

\author{J.\,Fink,$^{1,2,3}$ J.\,Nayak$,^2$ E.D.L.\,Rienks$,^{1,3}$ J.\,Bannies$,^2$ S.\,Wurmehl$,^{1,3}$ S.\,Aswartham$,^1$ I.\,Morozov$,^1$ R.\, Kappenberger$,^1$ 
 M.A.\,ElGhazali,$^{2,3}$ L.\,Craco,$^{1,4}$ H.\,Rosner,$^2$ C.\,Felser,$^2$  and B.\,B\"uchner$^{1,3}$}
\affiliation{
\mbox{$^1$Leibniz Institute for Solid State and Materials Research  Dresden, Helmholtzstr. 20, D-01069 Dresden, Germany}\\
\mbox{$^2$Max Planck Institute for Chemical Physics of Solids, D-01187 Dresden, Germany}\\
\mbox{$^3$Institut f\"ur Festk\"orperphysik,  Technische Universität Dresden, D-01062 Dresden, Germany}\\
\mbox{$^4$Instituto de Fisica, Universidade Federal de Mato Grosso, 78060-900 Cuiabá, MT, Brazil}\\
}

\date{\today}

\begin{abstract}
 In this Supplement we provide additional information on the evaluation of  scattering rates of the charge carriers  in LiFeAs. In addition we illustrate the difference of interband transitions in a normal Fermi liquid and a highly correlated electron liquid.
\end{abstract}

\pacs{74.25.Jb, 74.72.Ek, 79.60.−i }

\maketitle

\paragraph{Data evaluation.}

ARPES measures the energy ($E$) and momentum ($\mathbf{k}$) dependent spectral function $A(E ,\mathbf{k})$  multiplied by a transition matrix element and the Fermi function and convoluted with the energy and momentum resolution\,\cite{Damascelli2003}. The spectral function is given by 
\begin{eqnarray}
A(E,\mathbf{k})=\frac{1}{\pi}\frac{\frac{\Gamma(E,\mathbf{k})}{2}}
{[E -\epsilon _{\mathbf{k}}^*]^2+[\frac{\Gamma(E,\mathbf{k})}{2}]^2},
\label{26_1}
\end{eqnarray}
where $\Gamma$ is the lifetime broadening of the photo hole in the valence band  and  $\epsilon _{\mathbf{k}}^*$  is the renormalized particle energy.  $\epsilon _{\mathbf{k}}^*=\epsilon_{\mathbf{k}}- \Re\Sigma(E,\mathbf{k})$ and $\Gamma(E,\mathbf{k})=-2Z(E,\mathbf{k}) \Im\Sigma(E,\mathbf{k}) $. Here $\epsilon_k$ is the bare particle energy, $\Sigma$  is the is complex self-energy and $Z(E,\mathbf{k})$ is the renormalization function.
Strictly speaking the formula for the spectral function is only valid near the Fermi level and  for a small self-energy independent of momentum\,\cite{Mahan2000}. In this case the spectral function at  constant momentum has a maximum at $\epsilon_k^*$ and  a full width at half maximum of $\Gamma$. Very often the description of the spectral function is extended to higher energies\,\cite{Grimvall1981,Valla1999}.

 Normally the analysis of the  lifetime broadening is performed by taking slides through the measured intensity either at constant momentum or constant energy. The former method has the disadvantage that it is influenced by the Fermi function and an energy dependent background. Thus most of the studies use the latter method\,\cite{Damascelli2003}. In this case momentum distribution curves are fitted with Lorentzians yielding the momentum width. The energy width is then calculated by multiplying this momentum width  with the velocity.
 However also this method has drawbacks. In regions close to  the top or the bottom of  pockets the velocities are small and the momentum width is large.  In the present ARPES study of LiFeAs this occurs at the top of the inner and the middle of the hole  pockets and near the bottom of the inner electron pocket.  In those regions the finite energy and momentum resolution is of importance.
 
 Therefore we have analyzed our spectra by a multivariate fit of the spectra measured  in the two-dimensional $E-k$ space. We multiply the spectral function with the Fermi function and convolute it with the energy and momentum resolution. Waterfall plots of the energy and momentum dependent ARPES data together with a fit are presented in Figs.~1-5. The assignment of the data corresponds to Fig.~1(a) of the main text. The following parameters were derived: the energy dependent lifetime broadening $\Gamma(E)$ for all measured energies, a constant slightly asymmetric amplitude, values determining the dispersion, and values determining the slightly energy and momentum dependent background. In nearly all cases the fits are rather good. For the inner hole pocket (Fig.~5) the fit is less perfect probably due to small contributions to the spectral weight from the outer hole pocket.

\begin{figure} [H]
 \centering
\vspace{-3.0 cm}
\includegraphics[angle=0,width=1.4\linewidth]{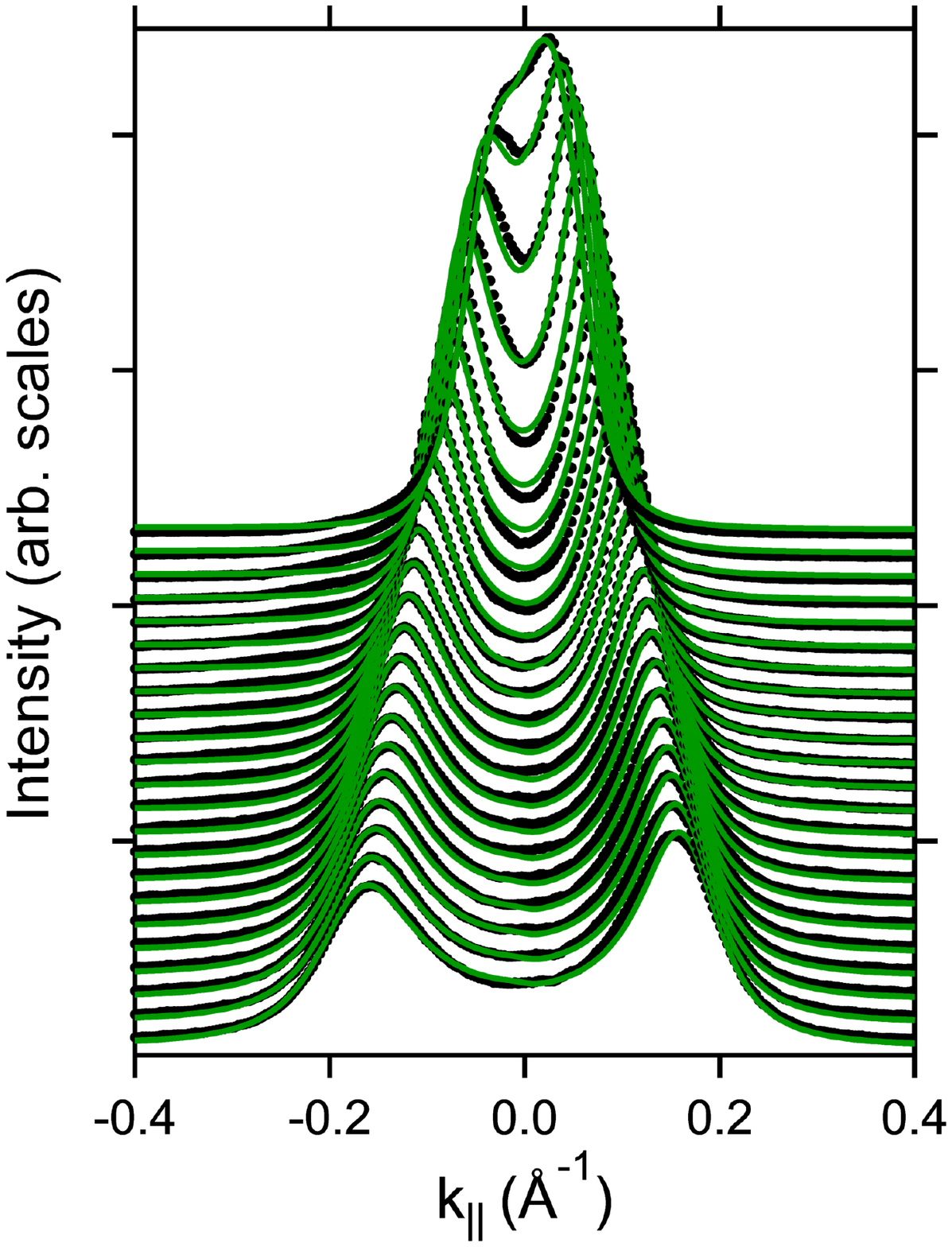}
\vspace{-1.0cm}
\caption{ Waterfall plot of  LiFeAs ARPES data (black dots) on the inner hole pocket [point (1)] together with a fit (green solid lines). The energy of the spectra range from 0.15 eV (lowest spectrum) to 0.02 eV (uppermost spectrum). The energy difference between the spectra is 0.006 eV.} 
\centering
\end{figure}
\begin{figure}[H]
 \centering
\includegraphics[angle=0,width=0.8\linewidth]{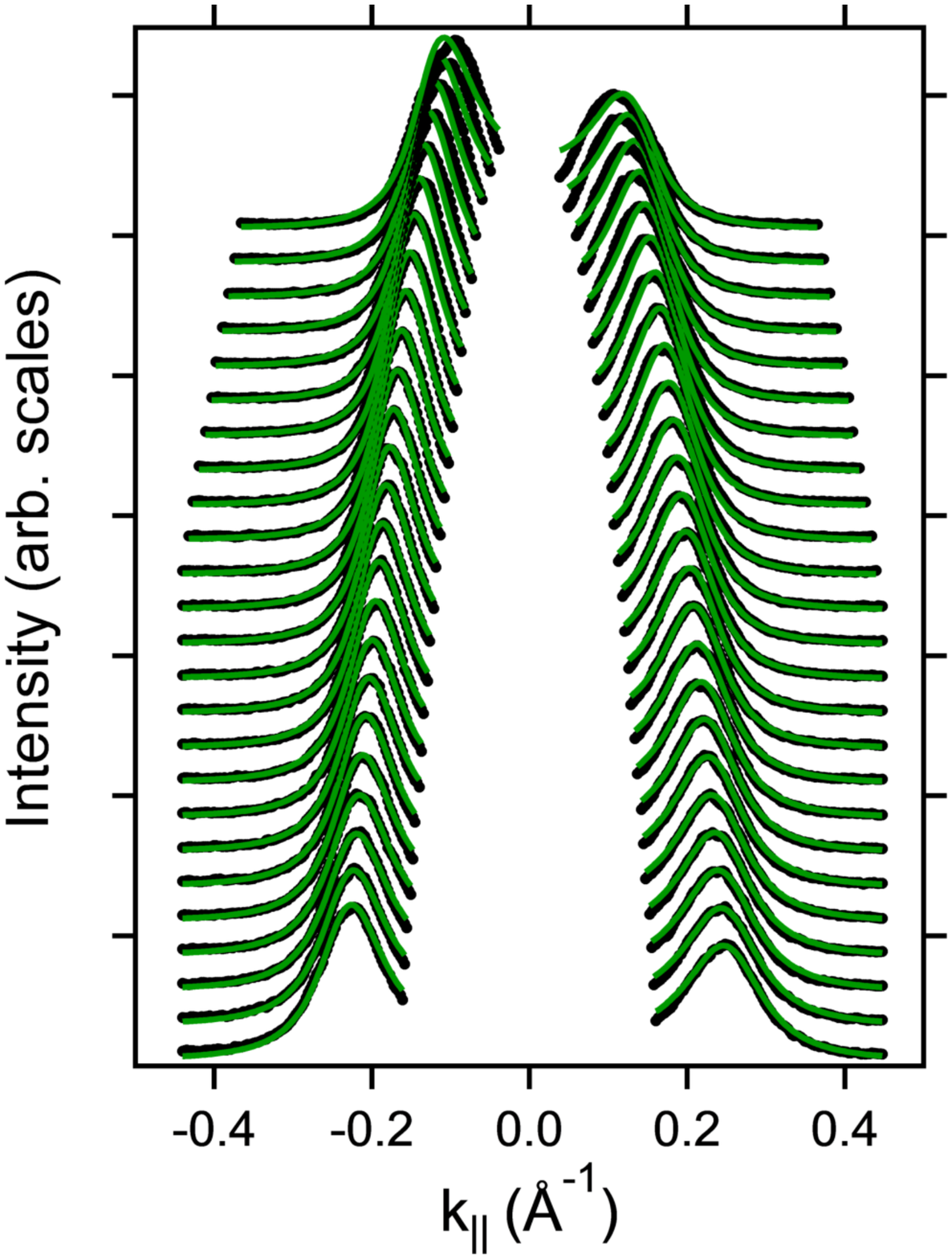}
\vspace{-2.0 cm}
\caption{(Color online)  Waterfall plot of  LiFeAs ARPES data  (black dots) on the middle hole pocket [point (2)] together with a fit (green lines). The energy of the spectra range from 0.0324 eV (lowest spectrum) to 0.0012 eV (uppermost spectrum). The energy difference between the spectra is 0.0024 eV. Contributions from the middle hole pocket are suppressed by a mask. } 
\centering
\end{figure}

\begin{figure}[H] 
 \centering
\includegraphics[angle=0,width=1.4\linewidth]{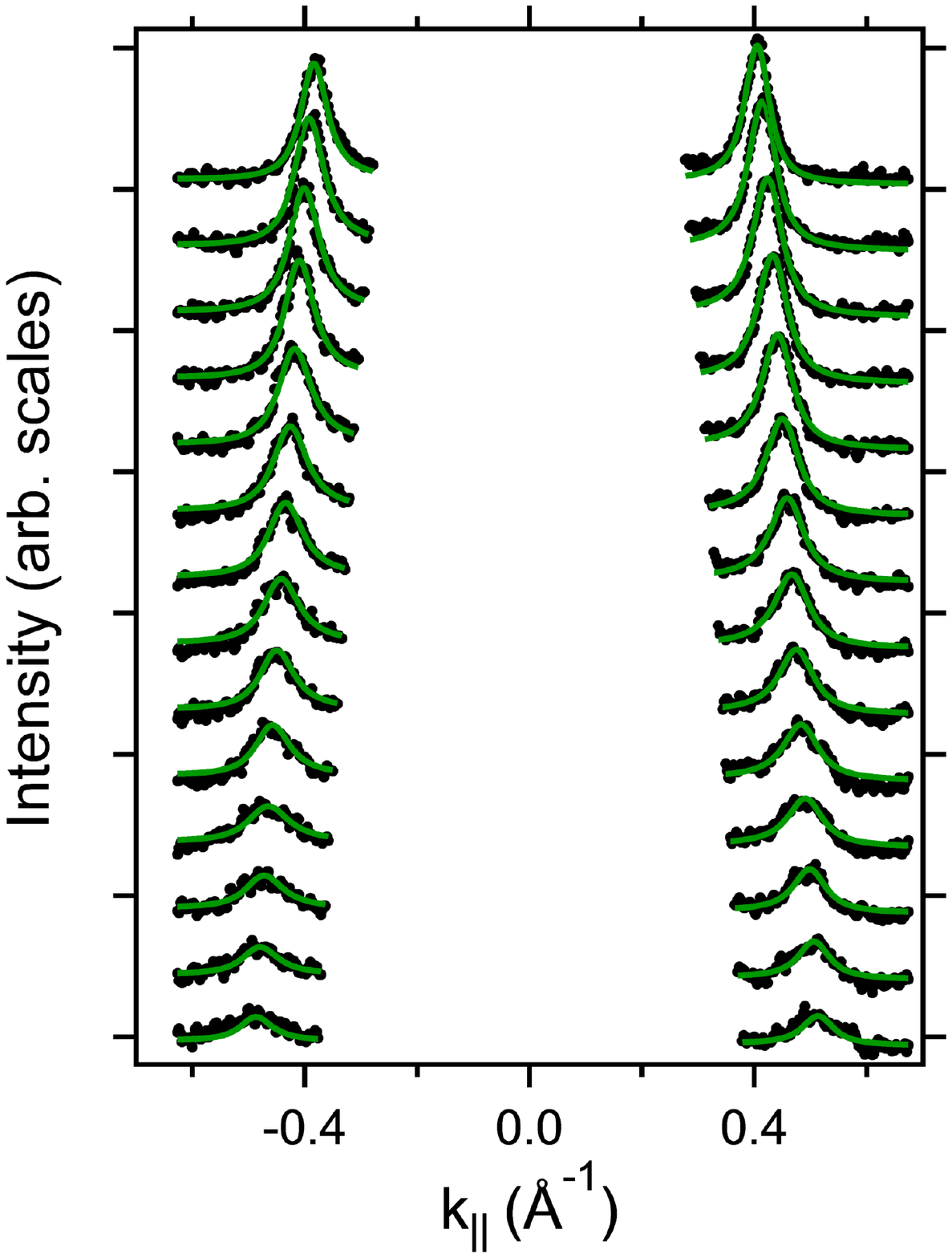}
%\vspace{-5.0 cm}
\caption{  Waterfall plot of  LiFeAs ARPES data (black dots)  on the outer hole pocket [point (3)] together with a fit (green solid  lines). The energy of the spectra range from 0.076 eV (lowest spectrum) to -0.01 eV (uppermost spectrum). The energy difference between the spectra is 0.006 eV. Contributions from the inner and the middle hole pocket are suppressed by a mask. } 
\centering
\end{figure}

\begin{figure} [H]
 \centering
\includegraphics[angle=0,width=1.6\linewidth]{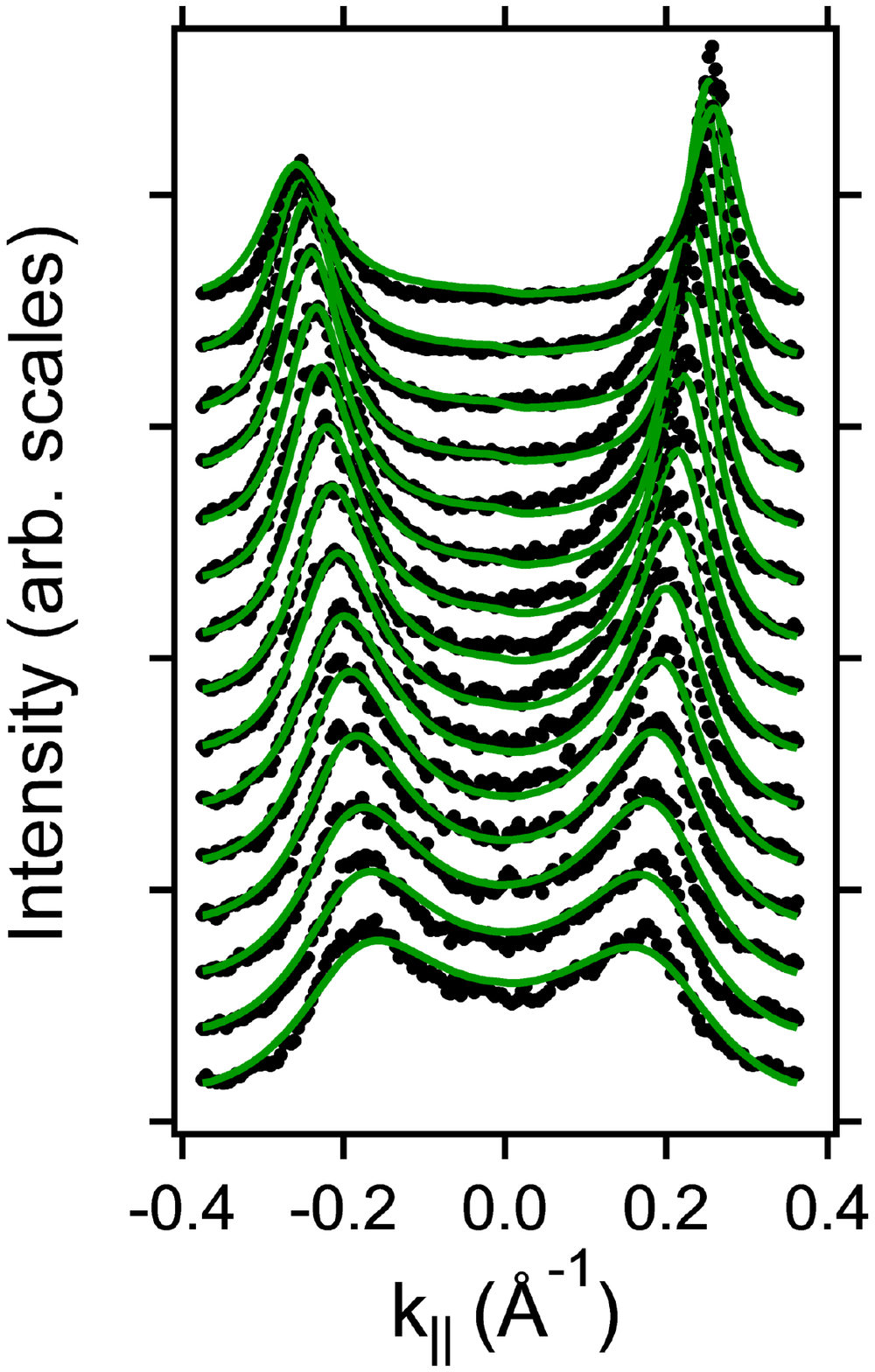}
\vspace{-1.0cm}
\caption{ Waterfall plot of  LiFeAs ARPES data (black dots) on the outer electron pocket [point (4)] together with a fit (green solid lines). The energy of the spectra range from 0.0465 eV (lowest spectrum) to 0.0045 eV (uppermost spectrum). The energy difference between the spectra is 0.003 eV. } 
\centering
\end{figure}

\begin{figure} [H]
 \centering
\includegraphics[angle=0,width=1.4\linewidth]{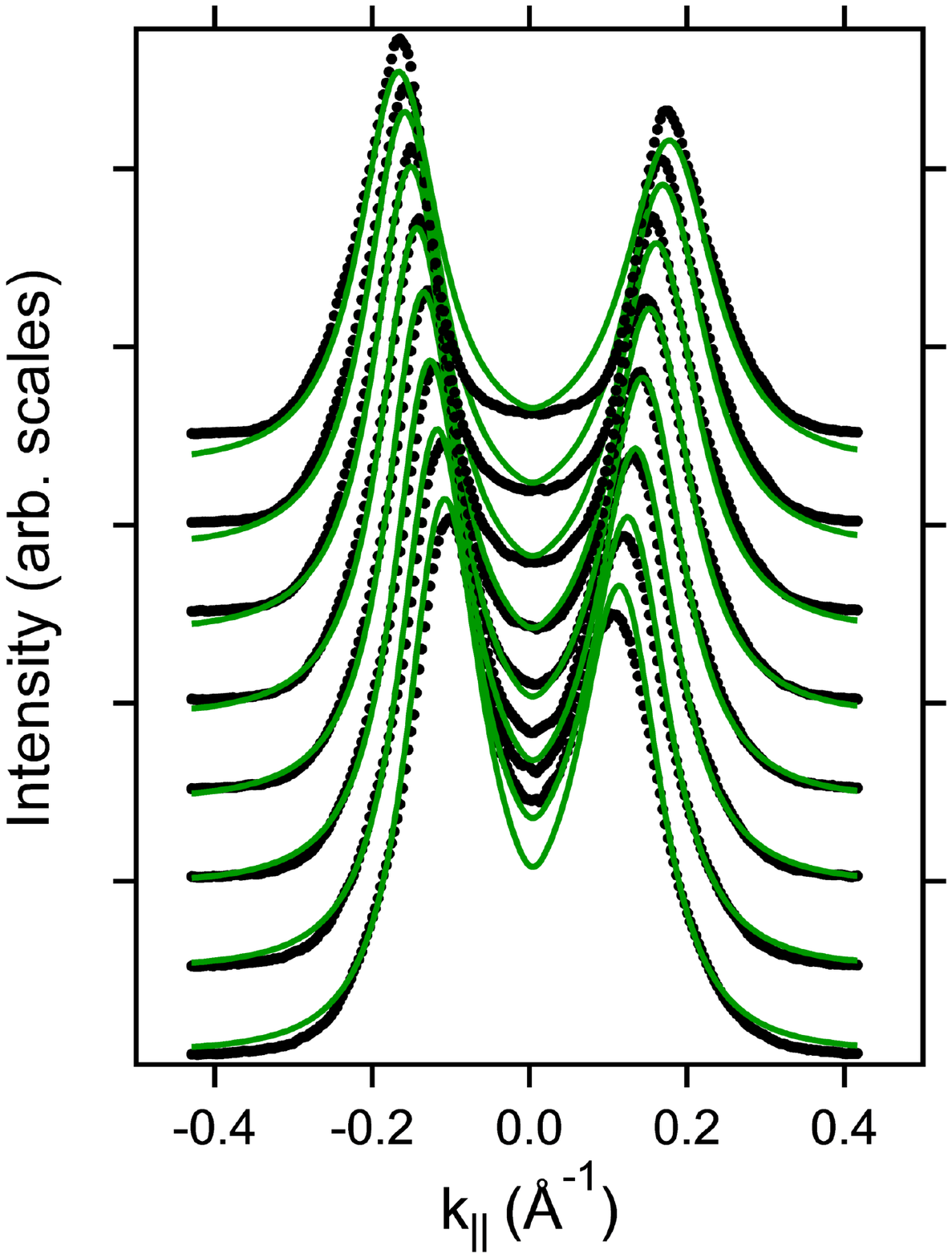}
\vspace{-1.0cm}
\caption{ Waterfall plot of  LiFeAs ARPES data (black dots) on the inner electron  pocket [point (5)] together with a fit (green solid  lines). The energy of the spectra range from 0.0285 eV (lowest spectrum) to 0.0075eV (uppermost spectrum). The energy difference between the spectra is 0.003 eV.  } 
\centering
\end{figure}

\begin{figure} [t]
%\hspace{-7 cm}
 \centering
%\vspace{-0.5cm}
\includegraphics[angle=0,width=1.6\linewidth]{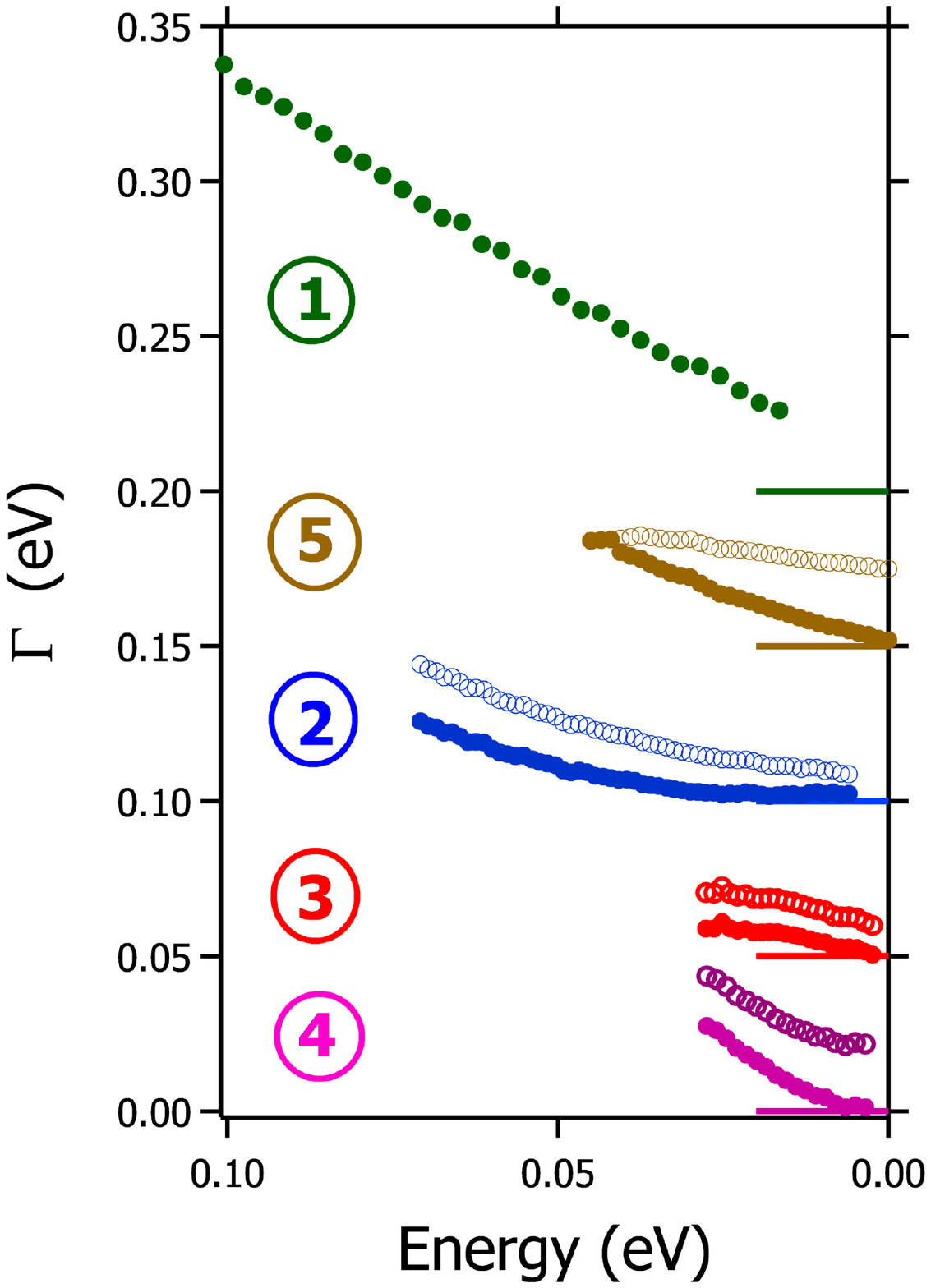}
%\vspace{-2 cm}
\caption{ Scattering rates of holes in LiFeAs at five different points in the Brillouin zone [see Fig.~1(a) of the main text]. Open circles: total scattering rates $\Gamma_t(E)$; closed circles: inelastic scattering rates $\Gamma_i(E)$. For point (1) we only show $\Gamma_i(E)$ because the corrections due to elastic scattering are small. } 
%\vspace{-6.0cm}
\centering
\end{figure}

In this way  we obtain the energy dependent total scattering rate  $\Gamma_t(E)$ which is the sum of the elastic scattering rate  $\Gamma_e(E)$  and the inelastic scattering rate $\Gamma_i(E)$. In highly correlated systems the latter is predominantly due to electron-electron interaction. To derive $\Gamma_i(E)$ one has to subtract the elastic from the total scattering rates. $\Gamma_e(E)$ is usually assumed to be constant
~\cite{Valla1999}.  However, this holds only for a linear dispersion when the velocity $v(E)$ is constant  because $\Gamma_e(E)= v(E)w_0$, where $w_0$ is the momentum width at the Fermi level related to the constant inverse mean free path between scattering sites inducing elastic scattering. To obtain the inelastic lifetime broadening for a parabolic dispersion, one has to subtract the energy dependent $\Gamma_e(E)$ from 
$\Gamma_t(E)$. In this context we mention that an energy dependent lifetime broadening on the basis of a constant momentum width has been discussed for  Cu surface states \,\cite{Kevan1983,Tersoff1983}. $w_0$  can be calculated from the data near $E_F$ because there $\Gamma_i(0)$ is zero and thus $\Gamma_e(0)=\Gamma_t(0)= v(0)w_0$; $v(E)$ is taken from the measured dispersion. This method cannot be used for those pockets which do not cross the Fermi level, i.e., the inner and the middle hole pocket. In this case we have used values from the $k_z=\pi$ plane where the middle hole pocket crosses the Fermi level. The $w_0$ values differ between different points. A theoretical explanation for this was presented recently from calculations of the elastic scattering rates at different sites of the Brillouin zone\,\cite{Herbig2016}.

In Fig.~6 we present the total scattering rates $\Gamma_t(E)$ together with $\Gamma_i(E)$. This figure clearly demonstrates that corrections of an energy dependent scattering rate is rather important for the strongly non-linear dispersions in LiFeAs.

\begin{figure} [t]
%\hspace{-7 cm}
 \centering
%\vspace{3.0cm}
\includegraphics[angle=90,width=1.2\linewidth]{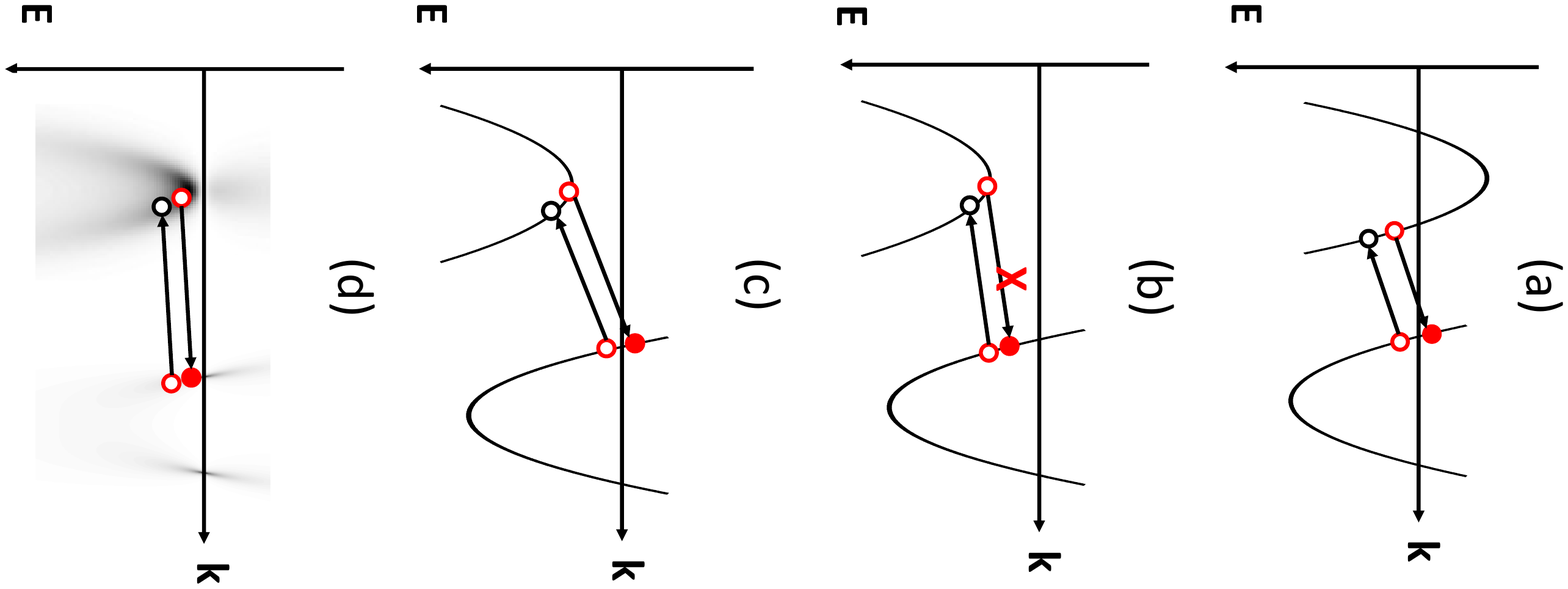}
\vspace{-1.0cm}
\caption{(Color online) Relaxation  process of a photo-hole in a hole pocket due to interband transitions between the hole pocket and an electron pocket. Black empty circle: initial photo hole. Open red circles: relaxed hole in the hole pocket and hole in the electron pocket. Filled circle: excited electron in the electron pocket. The black and the red symbols correspond to the initial and the final final state, respectively. (a) Transitions between an ungaped  hole pocket and an electron pocket for a weakly correlated Fermi liquid. (b) and (c) Analogous to (a) but for a gaped hole-pocket. (d) Analogous to (b) but for a highly correlated system.}
\centering
\end{figure}

\paragraph{Differences of the scattering rates between a weakly correlated and a strongly correlated system. }

In this section we describe Augerlike interband transitions between a hole pocket and an electron pocket which are related to the inelastic electron-electron scattering rate in hole pockets of LiFeAs. These transitions may also mediate antiferromagnetism and superconductivity in ferropnictides~\cite{Mazin2008a}. These transitions are illustrated in Fig.~7.
In Fig~7(a) we depict interband transitions in a weakly correlated system which lead to a relaxation of a hole, produced in the photoemission process, to lower energies. This leads in essence to a  mass enhancement. The photo-hole is filled by an interband transition from the electron pocket. The energy is transferred to an interband transition from a  state of the hole pocket at lower energies to unoccupied states of the electron pocket. Transitions into occupied electron pocket states are forbidden due to the Pauli principle. If the hole band is below $E_F$  [see Fig.~7(b) and (c)] transitions from the hole band to the electron band are only possible when the energy of the interband transition is larger than the energy of the top of the hole pocket $E_t$ [see Fig.~7(c)]. Thus at the top of the hole band  the scattering rate should go to zero because the phase space for the transition disappears. 

However, the situation  is completely different for a strongly correlated system. There the Fermi edge in momentum space is broadened leading to unoccupied states below the Fermi momentum. Therefore  interband transitions into states from the electron pocket below the Fermi level are allowed. The reason for this is that the Pauli principle in this case is no more operative. This leads to a finite phase space for interband transitions even at the top of the hole pocket and therefore to a finite scattering rate. This is illustrated in Fig.~7(d). Thus there is a fundamental difference of the scattering rates between  weakly correlated and strongly correlated systems. A similar argumentation was used to explain the enhanced scattering rates at low energies in highly correlated systems leading to the appearance of marginal Fermi liquids~\cite{Fink2016}.

\bibliographystyle{phaip}
\bibliography{Pnictide}